\documentstyle[twocolumn,prc,aps,epsfig]{revtex}
\newcommand{\be}{\begin{eqnarray}}
\newcommand{\ee}{\end{eqnarray}}

\def\<{\langle}
\def\>{\rangle}
\def\d{\partial}

\begin{document}

\twocolumn[\hsize\textwidth\columnwidth\hsize\csname @twocolumnfalse\endcsname
\title {\Large\bf
When can long-range charge fluctuations \\  serve as
 a QGP signal?
}
\author{
 E.V.~Shuryak$^1$ and M.A.~Stephanov$^2$
\bigskip
\\\small\it
$^1$ Department of Physics and Astronomy, State University of New York,
\\\small\it     Stony Brook, NY 11794-3800
\\\small\it
$^2$ Department of Physics, University of Illinois, Chicago, IL 60607-7059
\\\small\it and
\\\small\it RIKEN-BNL Research Center, Brookhaven National Laboratory, 
\\\small\it Upton, NY 11973
}
\date{\today}
\maketitle 
\begin{abstract}
We critically discuss recent suggestion to use long-range modes of
charge (electric or baryon) fluctuations as a signal for the presence
of Quark-Gluon Plasma at the early stages of a heavy ion collision.
We evaluate the rate of diffusion in rapidity for different
secondaries, and argue that for conditions of the SPS experiments, it
is strong enough to relax the magnitude of those fluctuations almost
to its equilibrium values, given by hadronic ``resonance gas''. 
  We evaluate the detector acceptance needed to measure
such ``primordial'' long-range fluctuations at RHIC conditions.  We
conclude with an application of the charge fluctuation analysis to the
search for the QCD critical point.
\end{abstract}
\vspace{0.1in}
]
\begin{narrowtext}   
\newpage 

\section{Introduction}

During the last few years 
significant attention has been attracted
to the issue of {\em event-by-event fluctuations} 
in high energy heavy ion collisions.
The original goal of such studies formulated in 
\cite{Stodolsky,Shu_fluct} has been focused on {\em 
equilibrium} thermodynamical
fluctuations
{\em at freeze-out}. It was argued that
long and intense final state interaction
of secondaries makes heavy ion collision very
different from hadronic collisions, in which dynamical fluctuations of
quantum origin produce quite different ``intermittent'' behavior.
Experimental data, such as obtained by NA49 collaboration \cite{NA49},
have indeed revealed only very small {\em Gaussian} fluctuations of different 
quantities 
incompatible with earlier predictions of large non-equilibrium
fluctuations, e.g. due to bubble formation during
the phase transition.  As discussed in \cite{SRS2,BH,DS} in detail,
extensive quantities like the total multiplicity, $N$, obtain 
contributions from {\em final state} interaction of secondaries
due to resonances, and
{\em initial state} fluctuations related to the number of
participant/spectator nucleons. Furthermore, experimentally, for central
collisions $\<(\Delta N)^2\>/\<N\>\approx 2$, and both effects share about
equally
the responsibility for moving this number away from the Poissonian
value  $\<(\Delta N)^2\>/\<N\>=1$. As emphasized in \cite{SRS2,koch1},
the intensive quantities, such as mean $p_T$, particles ratios
 etc. are determined
entirely by the  {\em final state} effects, and are well represented
by the resonance gas.
  
Other goals of the event-by-event fluctuation analysis
 have also been discussed. In particular,
one of us\cite{Shu_QM99}, using the analogy between Big Bang
phenomena
and the ``Little Bang'' in heavy ion collisions, suggested to
look for primordial ``frozen QGP oscillations'', as it is done for
 anisotropic components of microwave background in the Universe. 
The idea was inspired by an observation
that large-amplitude
oscillations can be excited due to  instabilities of 
parton counter-flows at the initial stage \cite{Mrow}.
On the other hand, two
simultaneous
papers, by Asakawa, Heinz and Muller\cite{AHM} and Jeon and
Koch\cite{koch2} suggested a different set of primordial signals:
{\em equilibrium} charge fluctuations in QGP, which happen
to be factor 2-3 {\em smaller} than in the hadron gas.

The idea \cite{AHM,koch2} is based on the well known
phenomenon of {\em kinetic slowdown} of fluctuation relaxation,
provided {\em sufficiently long-range harmonics} of a {\em conserved} 
charge density are considered.
If the relaxation time happens to be shorter
than the lifetime of hadronic stage of the collisions, 
the authors argue, the values of such fluctuations
should
deviate from their equilibrium (resonance gas) values towards 
their earlier, primordial values, typical for QGP.

Although this idea should work for parametrically long-range effects,
whether a number of necessary conditions can indeed be
fulfilled
in realistic heavy ion collisions depends on the answers
to the following questions: (i) How
fast is the relaxation due to final-state re-scattering in hadronic gas?
(ii) How large the 
detector acceptance should really be, in order
 to counter the relaxation? (iii) Is
there a window of parameters, forbidding resonance gas relaxation but
allowing for it at the QGP stage? (iv) If not, what fluctuations
should follow from non-equilibrium parton kinetics, at very early
pre-QGP stages of the collisions? 

In this paper we attempt to answer the first two of those questions.

\section{Relaxation of the long-range fluctuations}
\subsection{The setting}

In this section we  develop analytical
description of the time evolution of
fluctuations due to rescattering of charged particles in the hadronic
gas at the late stages of the heavy ion collision.\footnote{
Our approach is different from that of 
\cite{AHM,koch2,koch3,HJ,fialk,sasaki}.
We provide quantitative
description of the time evolution of long-range fluctuations based 
on a stochastic diffusion equation, which we solve analytically. 
The value of the diffusion
coefficient is our only numerical input and we determine it
using realistic hadronic gas properties.} The following
physical picture underlines our approach. 

We consider distributions
in rapidity space of a {\em conserved} net charge (e.g., electric or baryon).
We start our description from the beginning of the hadronic
phase. In each event this phase ``inherits'' a certain
distribution in rapidity space of the charge from 
the primordial quark-gluon plasma
phase. In the spirit of the statistical approach, we do 
not consider a specific charge distribution, but an ensemble
of distributions corresponding to a set of events.
This ensemble of charge distributions is
characterized by the mean and the magnitude of fluctuations. 
The initial magnitude of those fluctuations
is the input of our calculation. 

The fluctuations of different length, or range, in rapidity space (i.e.,
different harmonics of the charge distribution) relax on different
time scales. The fact that the net charge is conserved is crucial here.
Because the relaxation can only proceed via diffusion of the charge, 
the {\em longer} range fluctuations relax {\em slower}. The
relaxation time grows as a square of the range. Our goal
is to provide a quantitative description of this process.

In particular, we shall derive the equation which solves the
following problem: Given a certain detector rapidity acceptance
window, and given the initial magnitude of the charge fluctuations,
to find the time evolution of that magnitude. From the fact that
longer range charge fluctuations relax slower, it is evident that
fluctuations of total charge in a wider rapidity window relax slower.
Using typical numbers for the life-time of the hadronic phase
and rescattering properties
of the hadronic gas
we shall determine how wide the rapidity window must be chosen in
order to preserve, to a sufficient extent, the primordial (small) fluctuation
magnitude.

\subsection{Diffusion in rapidity space}

We introduce the density  of a conserved {\em net} charge $Q$ 
per unit rapidity,  $n(y,\tau)$,
as a function of time $\tau$ in a given event.
The objects we are 
really interested in are  the {\em fluctuations} of the density, 
\begin{equation}
f(y,\tau)=n(y,\tau) -  n_0,
\end{equation}
In the Bjorken boost invariant scenario
$n_0$ is proportional to the total charge in the central
region, $n_0=Q_{\rm total}/y_{\rm total}$, which is a constant
within each event.

We begin by
considering evolution of $f$ in a given event. 
The event-by-event fluctuations will be the subject of 
subsection \ref{sec:FP}. There we shall also address fluctuations
of the total charge ($Q$, or $n_0$), which we ignore for now.
Since $f$ is a density of a {\em conserved} quantity
it should obey a diffusion equation in
rapidity space:
\begin{equation}\label{f-diff}
\d_\tau f = \gamma\, \d^2_y f,
\end{equation}
where $\gamma=\gamma(\tau)$ is the diffusion coefficient.
This  diffusion equation (\ref{f-diff}) applies, provided the following
conditions are satisfied:\\
(i) 
We must assume that fluctuations
are sufficiently small, to justify linear approximation in (\ref{f-diff}).
This is reasonable if the number of particles
carrying the charge is large, which is fulfilled in sufficiently
central heavy ion collisions at the SPS and higher energies.\\
(ii) 
We must assume separation of scales --- the minimal interval of rapidity we
can consider must be much larger than the mean rapidity change of a charged
particle in a collision, $\delta y_{\rm coll}$. In other words,
$f$ must be averaged over sufficiently wide interval $\Delta y$.
As we shall see later, the typical $\delta y_{\rm coll}$ for
the electrical charge is of order 0.8, while for the baryon charge
it is 0.2. Even in such a large acceptance
 detector as STAR at RHIC, the coverage
does not exceed few units of rapidity. However, we can still consider a
somewhat idealized limit of rapidity windows much wider than
$\delta y_{\rm coll}$. This can
certainly be reasonable for the diffusion of the baryon charge.
Moreover, 
for the ideas we are evaluating to work, 
this limit is a necessary condition \cite{AHM,koch2}.
Indeed, if the rapidity window is not large compared to $\delta y_{\rm
coll}$, the fluctuations will relax by ballistic transport of the
charge on the time scale of the mean free time $\tau_{\rm free}$
towards their thermodynamic values, too fast for those ideas to work.
\footnote{In the same spirit, 
we shall not consider also the effects of finiteness
of $y_{\rm max}$, the total collision rapidity range. 
These have been studied already, \cite{koch3,HJ},
and correspond to certain boundary conditions in equation
(\ref{f-diff}).}

It will not be essential for the analysis of this section how the
coefficient $\gamma(\tau)$ is calculated. It may be (ideally)
determined from first principles of QCD.  In the late hadronic
stage of the collision, which is our concern in this paper,
one can also use a simple formula to estimate $\gamma$.
Since, on the microscopic
level, $\gamma$ is due to particle collisions:
\footnote{One can think of this as a random walk in rapidity space
with Gaussian random steps of
mean square length $(\delta y_{\rm coll})^2$ and compare the equation
for the mean square distance from origin: $(\Delta y)^2 = (\tau/\tau_{\rm
free}) (\delta y_{\rm coll})^2$ with the Gaussian solution of the
diffusion equation (\ref{f-diff}), 
which gives: $(\Delta y)^2 = 2\gamma \tau$.}
\begin{equation}\label{gamma}
\gamma = {(\delta y_{\rm coll})^2\over2\tau_{\rm free}}.
\end{equation}

Since we shall be using Eq.(\ref{gamma}) in Section
\ref{sec:realistic} to estimate $\gamma$ (or, rather, its time
integral, see below), we shall discuss it here.  It is clear that in
the collisionless limit, $\tau_{\rm free}\to\infty$, the diffision in
rapidity is absent, which is in trivial agreement with
Eq.(\ref{gamma}).  It is instructive to consider also the limit of
$\tau_{\rm free}\to 0$ --- the ideal hydrodynamic limit. In this limit
diffusion is also frozen --- particles have no time to
propagate. Eq. (\ref{gamma}) can still be applied, but one must
realize that, effectively, $\delta y_{\rm coll}$ also vanishes in this
limit. The reason is that successive rapidity shifts (each of
order unity) are not independent but strongly anticorrelated. This
happens because a colliding particle has no time to move out of the
region of particles of its initial rapidity, and is more likely to
scatter back in than further out. 
This is also related to the fact that space-time rapidity
$(1/2)\ln[(t+z)/(t-z)]$ and momentum rapidity can differ significantly
in this limit.

In this paper we need to describe late not-so-dense hadronic
stage of the collision with $\tau_{\rm free}$ only few times
smaller than evolution time, and we shall apply eq.(\ref{gamma})
without worrying about complications of the ideal hydro limit.  One
can also use cascade codes to estimate $\gamma$, or just
the diffusion length squared $2\int \gamma d\tau$, directly, with possible
anticorrelations already included, bypassing evaluation of $\delta
y_{\rm coll}$.

\subsection{Langevin equation}

Equation (\ref{f-diff}) describes relaxation of the density $n$
to its equilibrium value. It implies that we average over time
scales longer compared to the characteristic time of fluctuation
of $f$.
If, however, we are interested in
{\em fluctuations} of $n$ around its equilibrium value, equation
(\ref{f-diff}) is not enough. We must add a noise term:
\begin{equation}\label{f-langevin}
{\d_\tau f} = \gamma\, \d^2_y f + \xi(y,\tau).
\end{equation}
The characteristic autocorrelation time of the noise
 is of the same order as the time scale of the variation of $f$.
This time is very short since the rapidity windows we consider
contain many particles: $f$ changes each time any of the particles 
collides.

The rapidity correlation of the noise can be determined by
requiring that equilibrium fluctuations of $f$ are given
by thermodynamics. 
Thus, they are Gaussian with the probability distribution
\begin{equation}\label{Ptherm}
P_{\rm therm}[f] \sim \exp\left\{-\frac12\int dydy'\ {f(y)f(y')\over
2\,\chi(y-y')}\right\}, 
\end{equation}
where $\chi$ is the equilibrium 
density-density correlation function in rapidity,
which we need not know explicitly at this point%
\footnote{Relaxation time of small linear fluctuations we study
will not depend on their absolute magnitude.}%
,
and we used boost invariance of Bjorken scenario. In other words:
\begin{equation}\label{ffchi}
\<f(y)f(y')\> = \chi(y-y'),
\end{equation}
where the average is over time.
Writing the functional Fokker-Plank equation for the probability 
density $P[f]$ following from the Langevin equation 
(\ref{f-langevin}) we find that
(\ref{Ptherm}) is a stationary solution when the noise $\xi$ is
Gaussian and obeys:
\begin{equation}\label{noise}
\<\xi(y,\tau)\xi(y',\tau')\> = -2\gamma\,\d^2_y\chi(y-y')\,\delta(\tau-\tau'). 
\end{equation}

\subsection{Fokker-Plank equation}
\label{sec:FP}

Before we proceed to solving the Langevin equation (\ref{f-langevin}),
we should relate the time average in (\ref{ffchi}) to the
event-by-event average. They are the same by ergodicity.
However, one should also keep in mind that changes in initial
conditions will change the equilibrium value $n_0$. Such
initial state fluctuations, as discussed in the introduction,
can contribute significantly to the observed event-by-event
fluctuations of the {\em extensive} quantities, such as the
total charge $Q$. Such fluctuations are statistically
independent from the thermodynamic fluctuations we consider
and can be calculated separately and added in quadratures,
as is done in \cite{DS}. In contrast, for {\em intensive} quantities,
such as, for example, the ratio of positively to negatively
charged particles considered in \cite{koch1}, such initial
state fluctuations cancel and need not be considered.

Our goal now is to use equation (\ref{f-langevin}) to determine
how quickly fluctuations approach their thermodynamic distribution
(\ref{Ptherm}). In other words, we need to solve the functional
Fokker-Plank equation with a given initial distribution $P[f]$, and see how
this {\em functional} evolves with time.

The Fokker-Plank equation
corresponding to (\ref{f-langevin}), (\ref{noise}) is given by
\begin{eqnarray}\label{FP-y}
{\d_\tau P} = \int dy dy' \ 
\gamma\,{\delta\over\delta f(y)} &&\Big[
-\d^2_y  f(y) P 
\nonumber\\&&
-  \d^2_y  \chi(y-y')\,{\delta\over\delta f(y')} P \Big]\,,
\end{eqnarray}
and is not very easy to study. Boost invariance helps, however.
We shall do a Fourier
transform with respect to the variable $y$. Let us denote harmonics
of $f(y)$ by $f_k$.
If we start with the factorized probability distribution 
$P= \prod_k P_k[f_k]$, then each harmonic evolves independently
following partial differential equation
\begin{equation}\label{FP-k}
{\d_\tau P_k } =  \gamma\, k^2 {\d\over\d f_k} \left[
f_k P_k +  \chi_k {\d\over\d f_k}P_k \right].
\end{equation}
It is easier to derive this equation directly from the Fourier
transformed Langevin equation
\footnote{The condition of applicability of the diffusion
approximation, already discussed above, can be also stated
as the condition that we are describing sufficiently long-range,
i.e., small $k$ harmonics: $k<1/\delta y_{\rm coll}$. 
For high harmonics the relaxation
rate is $1/\tau_{\rm free}$ instead of $\gamma k^2$.}
\begin{eqnarray}
&&\d_\tau f_k = -\gamma\, k^2 f_k + \xi_k\,, 
\nonumber\\
&&\<\xi_k(\tau) \xi_{k'}(\tau')\> = 2\gamma k^2 \chi_k
\delta_{kk'}\delta(\tau-\tau')\,.
\end{eqnarray}

\subsection{Gaussian solution}

We can now pose the problem mathematically 
and solve it using equation (\ref{FP-k}).
Let us imagine that at some initial time $\tau=0$
(in our analysis -- the beginning of the hadronic phase)
 the probability
distribution for each harmonic $f_k$ is given by a Gaussian with
a mean square variance $\<f_k^2\>=\sigma_k^2$. What happens to
this probability distribution with time? Substituting the Gaussian
Ansatz
\begin{equation}
P_k[f_k] = (2\pi\sigma^2_k(\tau))^{-1/2} \exp\left\{-{
f_k^2\over 2\sigma^2_k(\tau)}\right\} ,
\end{equation}
one sees that
equation (\ref{FP-k}) preserves the Gaussian shape of the
probability distribution. The evolution of the distribution can
therefore be described by the evolution of $\sigma_k(\tau)$. The
equation for $\sigma_k(\tau)$ following from (\ref{FP-k})
is given by
\begin{equation}\label{dtausigma}
{\d_\tau}\sigma_k^2 = -2\gamma\, k^2 (\sigma_k^2 - \chi_k),
\end{equation}
with the solution:
\begin{equation}\label{sigma(tau)}
\sigma_k^2(\tau) = \chi_k + (\sigma_k^2(0) - \chi_k)
\exp\left\{-2 k^2 \int_0^\tau \gamma\,d\tau\right\}.
\end{equation}
For simplicity, we neglected the dependence of $\chi_k$ on $\tau$ in the
hadronic phase. Obviously, eq. (\ref{dtausigma})
can be solved for arbitrary $\chi_k(\tau)$, if one is given.
In this paper we apply eq. (\ref{dtausigma}) to
an idealized scenario, in which $\chi_k$
changes abruptly at the QGP-to-hadron-gas transition, and
see how the charge fluctuation magnitude relaxes to its hadron gas
equilibrium value. Eq. (\ref{sigma(tau)}) demonstrates explicitly
that, due to the local conservation of charge, harmonics with small $k$
relax very slowly. Using eq. (\ref{gamma}),
the exponent in (\ref{sigma(tau)}) can be also written as
$k^2 (\Delta y_{\rm diff})^2$, where
\begin{equation}\label{delta-diff}
(\Delta y_{\rm diff})^2 \equiv
2 \int_0^\tau \gamma\,d\tau
= \int_0^\tau (\delta y_{\rm coll})^2\, {d\tau\over\tau_{\rm
free}}\,
\end{equation}
is the mean distance
on which a charged particle diffuses during the time $\tau$.

\subsection{Relaxation in a rapidity window}

In experiment one is measuring the event-by-event 
fluctuation of charge, 
\begin{equation}
Q=\int_{y_1}^{y_2} n\,dy,
\end{equation}
in an interval of rapidity, $[y_1,y_2]$.
This fluctuation, $\Delta Q=Q-\<Q\>$,
 can be written as a weighted sum of the
fluctuations of the Fourier components $f_k$:
\begin{equation}\label{DeltaQck}
\< (\Delta Q)^2 \> = \int {dk\over2\pi}\ |c_k|^2\, \<f_k^2\> = 
\int {dk\over2\pi}\ c_k^2\, \sigma^2_k(\tau).
\end{equation}
The weight is determined by the Fourier harmonics, $c_k$, of the
function equal to unity  for $y\in[y_1,y_2]$ and zero
otherwise, i.e.,
\begin{equation}\label{ck}
|c_k|^2 = {\sin^2(k\Delta y/2)\over (k/2)^2}\ , 
\qquad (\Delta y\equiv y_2-y_1).
\end{equation}
The main contribution is from
harmonics with $k\sim 1/\Delta y$ and smaller. 
Equations (\ref{sigma(tau)}) and (\ref{DeltaQck}) determine the evolution
of the magnitude of the fluctuation $\< (\Delta Q)^2 \>$.

For example, consider the case when the
hadron gas ``inherits'' from the QGP phase
practically zero fluctuations of the charge (or fluctuations
much smaller than required by hadron gas thermodynamics ---
an idealization of \cite{AHM,koch2}), i.e., $\sigma^2_k(0)\ll\chi_k$. 
Then, using eqs. (\ref{sigma(tau)}),
(\ref{delta-diff}), (\ref{DeltaQck}), and (\ref{ck}), 
we can calculate the magnitude of the fluctuation of the charge $Q$
in the rapidity window $\Delta y$ after time $\tau$:
\begin{eqnarray}
\< (\Delta Q)^2 \> &=& \int {dk\over2\pi}\ 
{\sin^2(k\Delta y/2)\over (k/2)^2} \ 
\nonumber\\&&\times\ 
 \chi_k\left(1 - \exp\left\{-k^2 (\Delta y_{\rm diff})^2\right\}\right) 
\nonumber\\&&=
\chi_0 \Delta y\ F\left(x\right),
\nonumber\\&&
x\equiv {\Delta y_{\rm diff}\over(\Delta y/2)}.
\label{DeltaQ}
\end{eqnarray}
The dependence
on time $\tau$ and rapidity window $\Delta y$ enters through the ratio
$x$ of $\Delta y_{\rm diff}$ given by (\ref{delta-diff}) 
and half width of the window $\Delta y/2$.
\footnote{We assumed, for simplicity, 
that $\chi_k = {\rm const}$ in the interval
$k<1/\Delta y$, which is true when the density-density correlation length
is much smaller than $\Delta y$ --- a good approximation
for the large rapidity windows we consider: 
$\Delta y\gg\delta y_{\rm coll}$.} 
The function $F$ is given by:
\begin{eqnarray}\label{F}
F(x)= 1 + {x\over\sqrt{\pi}}\left(1-e^{-1/x^2}\right) - {\rm erf}
\left(1\over x\right).
\end{eqnarray}
It rises as $x/\sqrt{\pi}$ at small $x$ and saturates as
$1-1/(x\sqrt{\pi})$ at very large $x$ (Fig.\ref{fig:F}).
$\chi_0 \Delta y$ is the equilibrium thermodynamic size
of the fluctuation $\< (\Delta Q)^2 \>$ (it scales
linearly with the size $\Delta y$ as it should).
The reason that $F$ saturates only as a power of time, and not
exponentially, is the fact that characteristic relaxation
time of the long-range harmonics diverges.

\begin{figure}
\centerline{\epsfig{file=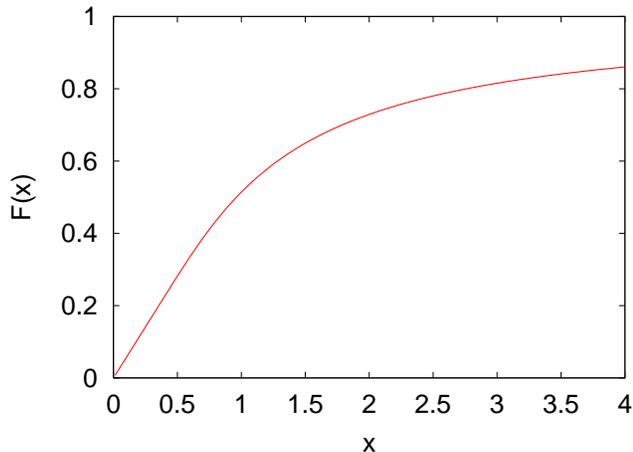,width=0.50\textwidth}}
\caption[]{
The function describing onset and saturation of the
equilibrium 
fluctuation magnitude in a rapidity window with time:   
(\ref{DeltaQ}), (\ref{F}).}
\label{fig:F}
\end{figure}

\section{Realistic rescattering in hadronic gas}
\label{sec:realistic}

\subsection{Cascade estimates of the diffusion rates}
Our estimates for $\Delta y_{\rm diff}$ in the hadronic gas phase can be
 simplified by noting that the dependence
of $\gamma$ on $\tau$ in (\ref{delta-diff}) is mostly due to
$\tau_{\rm free}$. The rapidity change per collision, $\delta y_{\rm coll}$,
is approximately independent of $\tau$, since the decrease of
the temperature is relatively weak in the whole hadronic stage,
$T=160-110$ MeV, and kinematics of low energy scattering
changes little.
In other words, $\Delta y_{\rm diff}$ is given by the random
walk formula
\begin{equation}
\Delta y_{\rm diff}=\delta y_{\rm coll}\ \sqrt{N_{\rm coll}},
\end{equation}
where 
\begin{equation}
N_{\rm coll}=\int_0^\tau {d\tau\over\tau_{\rm free}},
\end{equation}
is the total number of collisions per particle during the
time $\tau$, from the beginning of the hadronic phase, $\tau=0$,
until freeze-out.

At comparatively low relative energies of re-scattering in question,
all relevant processes are due to certain resonances, such as $\Delta$
for $\pi N$ collisions, $\rho$ for  $\pi\pi$ ones, etc. So, our first
task is to determine what is the mean change of rapidity in such
collisions. Taking pions and nucleons with thermal distribution, we
have found probabilities of different resonances and have determined
the
following values for the
mean rapidity shifts per collision
\be 
\delta y_{\rm coll}^\pi\approx 0.8; \qquad\delta y_{\rm coll}^N\approx 0.2.
\ee

The second (and much more involved) 
 task is to evaluate the average number of re-scatterings,
for each type of particles. The average value can be
evaluated from standard rates
\be N^i_{\rm coll}=\sum_j\int dt\ \sigma_{ij}\, v_{ij}(t)\, n_j(t) \ee
and even a very simple estimate shows that those numbers are not small.
The nucleons suffer especially multiple
collisions, of the order of 10-20,
since the $\Delta$-induced cross section at its peak is as large as
200 mb, and there are many pions around.
(This phenomenon is sometimes referred to as  {\em pion wind}.)
 However, more accurate estimates depend on many details, like the
 geometrical location
of the particle in question, expansion of the fireball etc.   Those
can
only be obtained from either realistic hydro simulations or from
hadronic cascades.

We have chosen the latter alternative, using 
two popular hadronic cascade codes. First, we have generated AuAu RHIC
events (100+100 GeV)
with RQMD\cite{RQMD}, and looked at the total rescattering
number per particle located at midrapidity $y\approx 0$.
 The results are shown in Fig.\ref{fig_RQMD}, as
a distribution over the number of rescatterings suffered by 
Kaons and Nucleons. One can see
that for  the
nucleons the distribution has a very wide tail, reaching large values
and with an average being of the order of 20. Kaons, on the other hand,
  suffer much smaller number of collisions, about 5 in average, and
there is about 20\% of kaons without rescattering at all. 

  Unfortunately standard RQMD output does not allow to trace an individual
pion, which appears and disappears, and we do not have the mean rescattering
pion number directly from that simulation.  We can only evaluate the
total number of $\pi-N$ rescatterings 
\be 
N_{\rm coll}^\pi= N_{\rm coll}^N \cdot{\<dN_N/dy\> \over \<dN_\pi/dy\>}\approx 2, 
\ee
but those are far from being dominant.

Fortunately,
a similar code, UrQMD \cite{UrQMD}, has been extensively studied,
and in the talk by M. Bleicher one finds a graph showing $dN_{\rm coll}/ds$
for
baryon-baryon, baryon-meson and meson-meson collisions, under the same 
conditions. Dividing by the number of mesons and integrating over
subcollision 
invariant energy, $s$, we obtained  the total 
number of collisions from meson-meson scattering to be
$N_{\rm coll}^{\rm meson}\approx 8 $  per meson. Most of the 
mesons are pions, of
course,
and this number fits well between N and K ones mentioned above.

\begin{figure}
\centerline{\epsfig{file=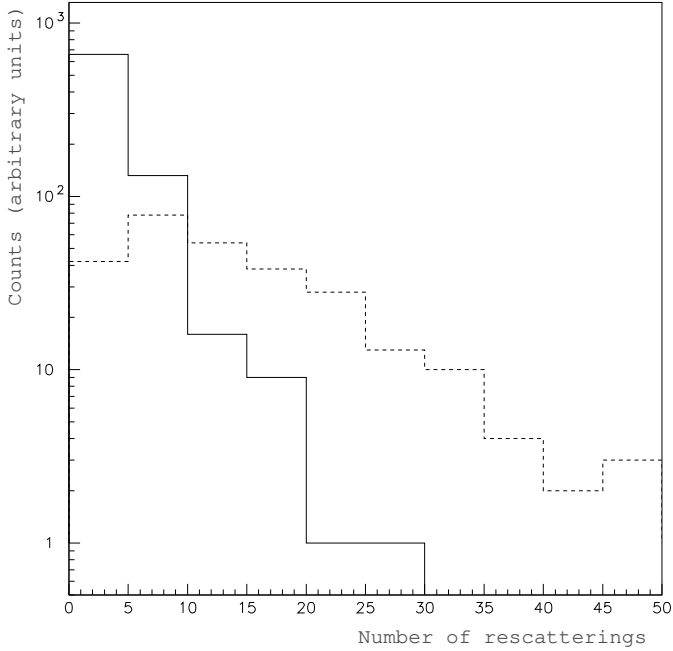,width=.55\textwidth}}
\caption[]{
\label{fig_RQMD}
Distribution over the number of rescatterings  in 
central AuAu 100+100
AGeV collisions, according to RQMD code. The solid line is for Kaons,
the dashed one for Nucleons, both taken in the mid-rapidity 
region ($|y|<2$).
}
\end{figure}


Adopting these numbers, we can combine it with
rapidity change per scattering, and get our final
estimates for the diffusion distance at RHIC:
\be\label{diff-rhic}
\Delta y_{\rm diff}^\pi\approx 2.2;\qquad \Delta y_{\rm diff}^N \approx 
0.9.
\ee

At SPS the pion multiplicity is about factor 2 smaller,
and the rescattering number is reduced to $N_{\rm coll}^N\approx 15$ and
$N_{\rm coll}^\pi\approx 5$, so that at
SPS
\be\label{diff-sps}
\Delta y_{\rm diff}^\pi\approx 1.8;\qquad \Delta y_{\rm diff}^N \approx 
0.77,
\ee
Let us now compare it to some typical experimental conditions
at SPS. For the 
accepted interval of rapidity of
$\Delta y=2$, the argument of $F(x)$ plotted in Fig.[1] is directly
these values of $\Delta y_{\rm diff}$, and one can read $F$ from it:
the result is 0.7 and 0.42 for pions and nucleons.


In the previous section we described an idealized case
$\sigma_k^2(0)\ll\chi_k$. For our estimate of the effect
we shall use more general expression, which gives the
time dependence of $\<(\Delta Q)^2\>$ given
its initial value at the beginning of the hadronic phase, $\tau=0$, 
$\<(\Delta Q)^2\>_0= \sigma_0^2(0)\Delta y$ \footnote{As with
$\chi_k$, we neglect
$k$-dependence in $\sigma_k$ in the interval of $k\sim 1/\Delta y$.}
and the hadronic gas equilibrium value $\<(\Delta Q)^2\>_{\rm
eq}=\chi_0\Delta y$:
\begin{eqnarray}\label{general}
\<(\Delta Q)^2\> &=&
\<(\Delta Q)^2\>_{\rm eq}\ F(x) + \<(\Delta Q)^2\>_0\ (1-F(x));
\nonumber\\&&
x = {\Delta y_{\rm diff}\over(\Delta y/2)}.
\end{eqnarray}

For example, starting from the fluctuations suppressed by a factor
of about 3 \cite{AHM,koch2}, i.e., $\<(\Delta Q)^2\>_0/\<(\Delta Q)^2\>_{\rm
eq}\approx 1/3$, and using $F\approx 0.70$ for the fluctuations of the electric
charge we find: $\<(\Delta Q)^2\>/\<(\Delta Q)^2\>_{\rm
eq}\approx 0.80$. I.e., only a 20\% suppression survives until freeze-out 
after a 3-fold suppression at the QGP-hadron-gas transition.
Similar estimate for the baryon number with $F\approx0.42$,
is more encouraging: $\<(\Delta B)^2\>/\<(\Delta B)^2\>_{\rm
eq}\approx 0.62$.

\subsection{Charge fluctuations at SPS}

 We have shown above, that the pion diffusion in
rapidity in the
hadronic phase is sufficiently strong.  
For a typical detector with the acceptance $\Delta y=2$
it reduces the initial deviation 
from equilibrium value of event-by-event charge fluctuation to nearly its
equilibrium value in the hadronic gas given by \cite{koch1}
\be \label{prelim}
{\<(\Delta Q)^2\> \over \<N_{\rm ch}\>}\approx 0.7,
\ee
which is very far from the QGP value. The deviation from the
  Poissonian value of 1 is due to the $\pi^+$, $\pi^-$ multiplicity
correlation induced by resonance decays.

Furthermore, a specific
 {\em centrality
dependence} of charge fluctuations in PbPb collisions 
follows from the resonance gas freezeout scenario.
The  magnitude of charge fluctuations (\ref{prelim})
should {\em increase}
towards
most central collisions.
This is opposite to what the QGP signal is supposed to do, if one
expects it to show up in more central collisions.

 This opposite trend is due to the fact that more central  collisions
correspond to {\em lower} freeze-out temperature \cite{HS}:  the
larger multiplicity is, the later freeze-out occurs. 
Significant centrality dependence of the radial flow observed at SPS
confirms this idea.
It follows, that for most central collisions one expects less
resonances, and correspondingly decreased deviation of charge
fluctuation from  the Poissonian value 1.

A simple estimate of this effect can be made using the fact
that the contribution of the resonances is controlled by the
Boltzmann factor. Thus, for example, a decrease of temperature
by $\Delta T\approx 20$ MeV leads to a decrease of
the contribution of the $\rho_0$ resonance by a factor of 2.
Similar decrease should occur in the contribution of 
other resonances, such as $\omega$. Therefore the variation
with centrality of the freezeout temperature of order $20$ MeV 
would correspond to centrality
dependence of (\ref{prelim}) of the order
of 10\%, rising towards more central events.

\subsection{Charge fluctuations at RHIC}

Finally, let us briefly discuss prospects of observation of QGP-like
charge fluctuations at RHIC. STAR detector has significantly larger
rapidity acceptance, and although diffusion of pions is slightly
stronger
at RHIC, one can hope to see deviations from the resonance gas value.
Indeed, for the rapidity window $\Delta y = 4$, equations (\ref{diff-rhic}),
(\ref{general}) predict 
that the primordial QGP factor 1/3 suppression 
of electric charge fluctuation will survive as factor 
\begin{equation}\label{DeltaQ-rhic}
{\<(\Delta Q)^2\>\over\<(\Delta Q)^2\>_{\rm eq}} \approx 0.7
\end{equation}
suppression. Another important prediction is that this suppression
must strengthen, i.e., the ratio (\ref{DeltaQ-rhic}) {\em decrease}, 
as the width $\Delta y$ is {\em increased}.
In other words, opening wider rapidity ``aperture'' allows to
see deeper back into the history of the collision.
\footnote{As emphasized in \cite{AHM,koch2,koch3,HJ}, $\Delta y$ should be
sufficiently smaller than the total collision rapidity range $y_{\rm
max}$. Otherwise the fluctuations will be trivially constrained by the total
charge conservation. In a sense, we would be looking beyond the
QGP stage of the collision.}

The baryon number fluctuations have slower diffusion, but rapid
proton-neutron
conversion together with
virtual invisibility of neutrons basically undermine the very idea:
 the relaxation slow-down can only occur if the current under consideration is
a conserved one. Similar problem arises if one considers
strangeness fluctuations. 

\section{$\pi^+/\pi^-$ fluctuations near the critical point $E$}

In this section we discuss a new signature of the critical end-point
$E$ \cite{SRS} based on the
$\pi^+/\pi^-$ fluctuations. As discussed in \cite{SRS,SRS2},
the main feature of thermodynamics near the critical point
is the presence of long-wavelength fluctuations of the
magnitude of the chiral condensate, or, the $\sigma$-field.
One of the signatures proposed in \cite{SRS,SRS2} was based
on the fact that such light $\sigma$-quanta cannot decay at freeze-out,
if it occurs near point $E$, and their large population
survives until the time after freeze-out when their rising mass exceeds
the 2$\pi$ threshold. The signature proposed in \cite{SRS,SRS2} used the fact
that produced pions had a necessarily non-thermal spectrum with
low mean $p_T$. 

The contribution of $\sigma$ decays
to the pion spectrum is, to some extent, 
 similar to contributions of other, standard, resonances in the
resonance gas model, such as $\rho^0$. 
In particular, the produced $\pi^+\pi^-$
pairs introduce positive correlation between the numbers
of positive and negative pions, therefore reducing the
fluctuation of $\pi^+/\pi^-$ ratio. The most significant
difference is that the contribution of $\sigma$ decays
 is only present near the critical point.

In order to estimate this contribution, we assume that the thermal population
of $\sigma$'s is approximately half of the so-called
``direct'' charged pions 
(since the mass of the sigma is comparable to $m_\pi$), 
as is done in \cite{SRS2}. Thus the decaying $\sigma$'s
produce the number of {\em charged} 
pions of order $2\times2/3\times1/2=2/3$ of the
number of ``direct'' charged pions. The latter is about 1/3 of the
total number, the rest are produced by resonance decays. Therefore,
the contribution of sigmas to the charge pion multiplicity is
about $2/3\times1/3\approx 20\%$. Thus, the contribution of $\sigma$'s
can be estimated to result in a reduction of fluctuations by an amount
of order 20\%. \footnote{
This can be compared to the
reduction due to $\rho^0$ resonance, which is about 9\% \cite{koch1}. 
Another 20\% in eq.(\ref{prelim}) is contributed
by $\omega$ and other heavier resonances.}

Even though the actual contribution will also depend
on the detector acceptance of the soft part of the pion spectrum,
it is reasonable to expect a noticeable suppression, comparable
to the suppression of about 30\% due to standard resonances ($\rho^0$,
$\omega$, etc.) in (\ref{prelim}).


What is important is that such an additional suppression
of the $\pi^+/\pi^-$ fluctuations can only appear if the
freeze-out occurs near the critical point.
As a function of the collision energy, this effect will be seen as
 a dip in the magnitude of the $\pi^+/\pi^-$ fluctuation, thus providing
a signature for the discovery of $E$. Away from the critical
point the fluctuations will be compatible with the
ordinary resonance gas result calculated in  \cite{koch1}.

The NA49 collaboration have now acquired data from collisions at 40, 80
and 160 AGeV of initial energy at SPS. We are looking forward
to the analysis of this data. If it shows a monotonic dependence of
$\pi^+/\pi^-$ ratio fluctuation on collision energy as well as
on centrality, consistent with the resonance gas
prediction, then one should be able to rule out
the presence of the critical point in regions
of the QCD phase diagram the location of which 
can be determined by the analysis similar to \cite{pj-PbPb}.
Collisions at other energies, in particular those of RHIC,
will be needed to scan other regions of the phase diagram.
If a signal described here is observed, other signatures
discussed in \cite{SRS2} must also be seen to confirm the
discovery of the critical point.


\section*{Acknowledgement} Authors are grateful to D. Son and
T. Schaefer for discussions. ES is partly supported by  the US DOE grant
No. DE-FG02-88ER40388.

\end{narrowtext}
\end{document}